# Teaching with Code: Globular Cluster Distance Lab
## James Newland
### Bellaire High School, Houston ISD, Houston, Texas

## Teaching with Code: Globular Cluster Distance Lab

This interactive activity is meant to let students engage with real astronomical data in the context of coding. Students will use code to perform photometric analysis on several images from a target variable star. Background information about variables stars, luminosity, the inverse square law, standard candles, and distance modulus follows the common Astro 101 curriculum. This way, the activity can act as an introduction or augment existing material. Links to even more information are available so students can be sure they have the background material down before pressing on to photometry or coding. Note that the activity uses the classic magnitude scale such that luminosities must be converted to magnitudes and vice versa. The code for switching between the two systems is provided. The original idea for this exercise came from an aperture photometry project with no coding. All of the images used for this activity were taken using the Skynet Robotic Telescope Network by the author.[6]

## Student Coding Exercises

How much help students need with the coding exercises depends on their comfort with computer programming. This sort of activity works well with pair programming but can also be done solo. The project can be done in a lab environment or virtually. This project was designed using the popular Jupyter notebook platform running Python and can run locally or in the cloud.[3] Unfinished code blocks require the student to combine existing functions and some algebra to produce results. The comments guide the student in shaping a solution to produce the desired output. Ellipses in the code indicate the student must complete a particular section. The distance_modulus function asks a student to code the well-known astronomical relation to find the distance to an object, if the apparent and absolute magnitude is known. Although a known value is provided using the absolute and apparent magnitude of the sun to allow the student to test the code, this is far from a robust testing protocol. The depth can be altered by the instructor if desired.

Next, students work through the process_image function using comments as a guide. This function will determine the apparent magnitude of our target RR Lyrae star. Note the use of function composition, which may be new for students. Each component expects students to investigate the particular code involved and to figure out how it fits into the bigger picture. This function will be used to process all of our images. Next, students use the given parameters in the comments to make the process_image function analyze each image to determine the apparent magnitude of the target star. The images were taken throughout the night, and the brightness's variation should be evident as the student processed the image set. The various apparent magnitudes are added to a list as we work through the images to make the comparison more straightforward. The code produces images showing the extracted sources and the magnitude of the target star and calibration star.

Finally, the student can use the results of the photometry and the distance modulus function to determine the distance to the host star cluster. The coding style was meant to emulate the sort of programming tasks astronomers often have to complete. The use of common packages, functions, and data structures means this project has the look and feel of a typical data reduction task.

**Learning Data Reduction by Example**

Many Astro 101 students have very little exposure to techniques in data reduction. The aperture photometry technique used here does more than find sources and measure the flux. The detection and subtraction of the background from the data allow students to get a sense of the signal to noise ratio. The images shown in the notebook display the original image data with the detected background shown in the middle and the resulting cluster image with the background subtracted out. Background subtraction used the SEP and SExtractor packages.[1,2]

Having the complete function allows students to explore the actual process of background subtraction. An instructor could also remove parts of the code and have advanced students make it work as intended.

The algorithm for extracting sources and highlighting them in is shared in the code. This algorithm itself is modified from the documentation for the given package. This code is meant to demonstrate the data reduction process and is built from existing examples taken from literature. This sort of information sharing and dependency on other researchers shows students an example of the scientific collaboration found in the sciences. Again, some parts of this function could be removed to make the exercise more rigorous.

**Computing Pedagogy in Astronomy**

Computational thinking is increasingly a part of the practice of science. Science educators are responsible for teaching student's science knowledge but also science as a discipline. This lesson uses computational thinking in more than the sense of computer programming. Some students could engage with the lesson and not fully understand the computer science paradigms behind the algorithms. Other students could engage with the lesson and explore the deeper computer programming paradigms like iteration, Boolean logic, and data structures. Future work will involve a more thorough attempt to combine known computer science teaching pedagogy with science teaching pedagogy.

This lesson was first published via TeachEngineering.org. The project requires some basic knowledge of Python and Jupyter notebook. Besides the published lesson, the individual Jupyter notebook files are available for students with the code left out and for teachers with the code completed.
https://www.teachengineering.org/makerchallenges/view/rice3-2467-milky-way-stars-design-challenge

The Jupyter notebooks were also tested in two separate cloud computing platforms, Microsoft Azure Notebooks & Google Colab.

Links to just the notebooks themselves can be found at
https://www.jimmynewland.com/wp/astro/measuring-the-milky-way-with-stars/

When creating this exercise, the astronomy Python community was invaluable. Another lesson for students is how many people a single researcher depends on to complete a given task. A special mention goes to Astrobetter, Python4Astronomers, and the AstroPython community. Extensive use of AstroPy and NumPy were critical for creating this activity.[4,5,7]

This project was developed through a combination of two NSF grant-funded programs: Teacher research internship at McDonald Observatory under the direction of University of Texas at Austin Department of Astronomy (NSF Grant No. AST-1616040 principal investigators Dr. C. Sneden & Dr. K. Finkelstein) and Expeditions in Computing Research Experience for Teachers under the direction of the Department of Electrical and Computer Engineering and the Office of STEM Engagement at Rice University (NSF Grant No. IIS-1730574 principal investigators Dr. A. Sabharwal & Dr. C. Nichol). Thanks also to Christina Crawford and Allen Antoine from the Rice Office of STEM Engagement. And thanks to Asa Stahl from Rice astrophysics and my colleague Justin Hickey from Episcopal High School in Bellaire, Texas, for inspiring me to create a Python astronomy lesson and helping with feedback.